\documentclass{elsart}
\usepackage{epsfig}
\usepackage{bm}
\usepackage{graphicx}

\newcommand{\rt}{\rightarrow}
\newcommand{\bds}{\bar{B^0} \rt D^{*+}\ell^-\bar{\nu}}
\newcommand{\bdse}{\bar{B^0} \rt D^{*+}e^-{\bar{\nu}}}
\newcommand{\bdsm}{\bar{B^0} \rt D^{*+}\mu^-\bar{\nu}}
\newcommand{\dl}{D^{*+}e^-}
\newcommand{\dst}{D^{*+}}
\newcommand{\ds}{D^{*+} \rt D^0\pi^+}
\newcommand{\da}{D^0 \rt K^- \pi^+}

\newcommand{\vcb}{\left|V_{cb}\right|}

\newcommand{\fb}{\ensuremath{\rm fb^{-1}}}
\newcommand{\rha}{\rho^2_{A_1}}
\newcommand{\rhf}{\rho^2_{F}}
\newcommand{\umon}{{\rm{GeV}}/c}
\newcommand{\umonm}{{\rm{MeV}}/c}
\newcommand{\umass}{{\rm{GeV}}/c^2}
\newcommand{\umassm}{{\rm{MeV}}/c^2}
\newcommand{\umasss}{{\rm{GeV}^2}/c^4}

\newcommand{\vfca}{3.54}
\newcommand{\vfcb}{0.19}
\newcommand{\vfcc}{0.18}

\newcommand{\rhca}{1.35}
\newcommand{\rhcb}{0.17}
\newcommand{\rhcc}{0.19}

\newcommand{\brca}{4.59}
\newcommand{\brcb}{0.23}
\newcommand{\brcc}{0.40}
\newcommand{\vf}{(\vfca\pm\vfcb\pm\vfcc)\times 10^{-2}}
\newcommand{\rh}{\rhca\pm\rhcb\pm\rhcc}
\newcommand{\br}{(\brca\pm\brcb\pm\brcc)\times 10^{-2}}

\newcommand{\vcbf}{(3.88\pm0.21\pm0.20\pm0.19)\times10^{-2}}
\newcommand{\vfst}{(\vfca\pm\vfcb)\times 10^{-2}}
\newcommand{\rhst}{\rhca\pm\rhcb}
\newcommand{\brst}{\brca\pm\brcb}

\begin{document}

\begin{frontmatter}

\begin{flushleft}\includegraphics[width=3.5cm]{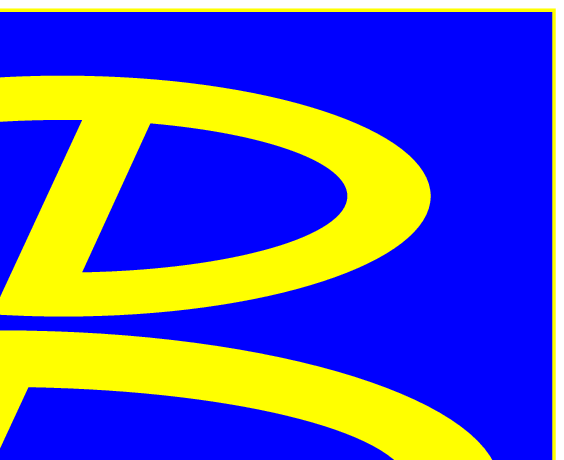}\end{flushleft}
\vspace{-2.5cm}
\hspace{8.5cm}
{\tt KEK preprint 2001-149}\\
\hspace{8.75cm}
{\tt Belle preprint 2001-19}

\vspace{2cm}

\title{Determination of $\vcb$ using the semileptonic decay $\bdse$}


\author{Belle Collaboration} 

\begin{center}
{\normalsize
  K.~Abe$^{9}$,               
  K.~Abe$^{43}$,              
  R.~Abe$^{32}$,              
  T.~Abe$^{44}$,              
  I.~Adachi$^{9}$,            
  Byoung~Sup~Ahn$^{17}$,      
  H.~Aihara$^{45}$,           
  M.~Akatsu$^{25}$,           
  Y.~Asano$^{50}$,            
  T.~Aso$^{49}$
% ...
}\end{center}

\address{
% ...
$^{9}${High Energy Accelerator Research Organization (KEK), Tsukuba}\\
% ...
$^{17}${Korea University, Seoul}\\
% ...
$^{25}${Nagoya University, Nagoya}\\
% ...
$^{32}${Niigata University, Niigata}\\
% ...
$^{43}${Tohoku Gakuin University, Tagajo}\\
$^{44}${Tohoku University, Sendai}\\
$^{45}${University of Tokyo, Tokyo}\\
% ...
$^{49}${Toyama National College of Maritime Technology, Toyama}\\
$^{50}${University of Tsukuba, Tsukuba}\\
% ...
}

\normalsize

\begin{abstract}

We present a measurement of the Cabibbo-Kobayashi-Maskawa (CKM) matrix 
element $|V_{cb}|$ using a 10.2 $\rm{fb^{-1}}$ data sample 
recorded at the $\Upsilon(4\rm{S})$ resonance with the Belle detector 
at the KEKB asymmetric $e^+e^-$ storage ring.
By extrapolating the differential decay width of the $\bdse$ decay to the 
kinematic limit at which the $D^{*+}$ is at rest with respect to the 
$\bar{B^0}$, we extract the product of $|V_{cb}|$ with the normalization of 
the decay form factor $F(1)$, $\vcb F(1)$ = $\vf$, 
where the first error is statistical and the second is systematic.
A value of $|V_{cb}|$ = $\vcbf$ is obtained using a theoretical calculation
of $F(1)$, where the third error is due to the theoretical uncertainty in the
value of $F(1)$.
The branching fraction ${\mathcal{B}}(\bdse)$ is measured to be $\br$.

\vspace{3\parskip}
\end{abstract}

\end{frontmatter}
\clearpage


\section{Introduction}

Accurate measurements of CKM matrix elements are important to constrain
the Standard Model. Knowledge of $\vcb$ is needed to relate a number
of experimentally measured $\it CP$ violating observables to the 
Wolfenstein parameters~\cite{Wolf} through which they are usually illustrated.
$\vcb$ can be determined by studying exclusive
$\bds$ decays using Heavy Quark Effective Theory (HQET)~\cite{HQET},
an exact theory in the limit of infinite quark masses.
The development of HQET yields an  expression for the 
$\bds$ decay rate in terms of a single unknown form factor. 
The form factor is parameterized in terms of ${\rm y}$, 
the inner product of the $\bar{B^0}$ and $D^{*+}$ four-velocities:

\[
{\rm y} = v_{\bar{B^0}} \cdot v_{D^{*+}} = 
\frac{M^2_{\bar{B^0}} + M^2_{D^{*+}} - q^2}{2M_{\bar{B^0}}M_{D^{*+}}},
\]

\noindent
where $M_{\bar{B^0}}$ and $M_{D^{*+}}$ are masses of the $\bar{B^0}$ and
$D^{*+}$ meson respectively and
$q^2 = ( p_{\bar{B^0}} - p_{D^{*+}} )^2$.
Here, and in what follows,
$p$ stands for the four-momentum vector of the particle in subscript.

The form factor $F({\rm y})$ is expressed as the product of a normalization 
factor $F(1)$ and a shape function which is constrained by a dispersion
relation~\cite{disper_1998}.
We extract the product of $\vcb$ and $F(1)$ from extrapolation of the
data to the zero-recoil point ${\rm y} = 1$, 
where the $D^{*+}$ is at rest with respect to the $\bar{B}^0$.

At this kinematic point, HQET allows us to calculate $F(1)$ with
small and controllable theoretical errors.
There are several corrections to the heavy quark limit $F(1) = 1$  as follows:

\[
F(1) = \eta_{\rm{QED}}\eta_A ( 1 + \delta_{1/m^2} + \ldots ).
\]

\noindent
QED correction factors up to leading logarithmic order provide 
$\eta_{\rm{QED}} \approx 1.007$~\cite{sirlin}.
$\eta_A$ is a short-distance correction arising from the finite QCD
renormalization of the flavor-changing axial current at the point of zero 
recoil.
The first order term in the non-perturbative expansion in powers of
$1/m_Q$ vanishes by virtue of Luke's theorem~\cite{luke}, and 
$\delta_{1/m^2}$ represents the second order correction.
By contrast, the form factor of $\bar{B} \rt D \ell \bar{\nu}$ decay is not
protectected against $1/m_Q$ corrections at zero recoil~\cite{dlnu_mq}, 
hence the comparison of results from $\bar{B} \rt D^* \ell \bar{\nu}$
with $\bar{B} \rt D \ell \bar{\nu}$~\cite{dlnu1}-\cite{dlnu_belle} 
provides some measure of the size of corrections to HQET.
Results for $\eta_{\rm{QED}}$ and $\eta_A$ are combined with their errors
added linearly to give $F(1) = 0.913\pm0.042$~\cite{Babar}.
This value of $F(1)$ is in good agreement with the latest lattice QCD 
calculation, $F(1) = 0.9130^{+0.0293}_{-0.0348}$~\cite{lattice_F1}.

This method currently gives the smallest theoretical error
in the extraction of $\vcb$.
Measurements of $\vcb$ based on the dispersion relation have been published
by the OPAL~\cite{vcb_opal} and DELPHI~\cite{vcb_delphi2} collaborations 
using $B$ mesons produced in $Z^0$ decays at LEP and reported by
the CLEO collaboration~\cite{vcb_cleo2} operating at the 
$\Upsilon(4\rm{S})$ resonance.

In this paper, we present a measurement of the branching fraction
of the exclusive semileptonic $B$ decay 
$\bar{B^0} \rt D^{*+} e^- \bar{\nu}$, and a determination
of $\vcb$. The charge conjugate mode is implicitly included.
For this precise measurement we use $\bdse$ and not $\bdsm$ since
the electron identification efficiency plateau has a wider range in 
both momentum and polar angle than that of muons, leading to a better
control of systematic uncertainties.

The data correspond to an integrated luminosity of 
$10.2$ fb$^{-1}$ accumulated at the $\Upsilon(4\rm{S})$ resonance 
with the Belle detector~\cite{NIM} at KEKB~\cite{kekb}, an asymmetric $e^+e^-$ 
storage ring.
This corresponds to $10.8\times 10^6$  $B\bar{B}$ events.
An additional sample with an integrated luminosity of $0.6 $ fb$^{-1}$
taken at an energy 60 MeV below the $\Upsilon(4S)$ resonance is used as a
control sample to check continuum processes.
A GEANT~\cite{geant} based Monte Carlo (MC) simulation is used to study 
the signal decay mode and also to estimate some backgrounds.


\section{Belle Detector}

Belle is a general-purpose detector which includes a 1.5~T superconducting
solenoid magnet.
Charged particle tracking is provided by a silicon vertex detector (SVD) 
and a central drift chamber (CDC) that surround the interaction region. 
The SVD consists of three layers of double-sided silicon strip detectors;
one side of each detector measures the $z$ coordinate and the
other the $r-\phi$ coordinate. 
The CDC has  50 cylindrical layers of anode
wires; the inner three layers have instrumented cathodes
for $z$ coordinate measurements. 
The charged particle acceptance covers laboratory polar angles between
$\theta=17^\circ$ and $150^\circ$ corresponding to about 92\%
of the full solid angle in the center of mass (CM) frame. 
The momentum resolution is 
$(\sigma_{\left|\bm{p}_t\right|}/\left|\bm{p}_t\right|)^2 = 
(0.0019\left|\bm{p}_t\right|)^2 + (0.0030)^2$,
where $\bm{p}_t$ is the transverse momentum in the laboratory frame, 
in units of $\umon$. 

Charged hadron identification is provided by $dE/dx$ measurements in
the CDC,  a mosaic of 1188 aerogel \v{C}erenkov counters (ACC), and a 
barrel-like array of 128 time-of-flight scintillation counters (TOF). 
The $dE/dx$ measurements have a resolution for hadron tracks of 6.9\% and 
are useful for $\pi /K$ separation for $\left|\bm{p}_{\rm lab}\right| < 
0.8$~$\umon$ and $\left|\bm{p}_{\rm lab}\right|> 2.5$~$\umon$.
Here, and in what follows, $\bm{p}_{\rm lab}$ is the three-momentum 
vector in the laboratory frame and $\bm{p}$ denotes the three-momentum 
vector in the CM frame.
The TOF system has a time resolution for hadrons of $\sigma \simeq 100$~ps 
and provides $\pi /K$ separation for  $\left|\bm{p}_{\rm lab}\right| 
< 1.2$~$\umon$.~  
The ACC covers the range $1.2~\umon < \left|\bm{p}_{\rm lab}\right| 
< 3.5~\umon$ and the refractive indices of the ACC elements vary with
polar angle to match the kinematics of the
asymmetric energy environment of Belle.
Particle identification probabilities are determined from
the combined response of the three systems.  

The electromagnetic calorimeter (ECL) consists of an array of 
8736 CsI(Tl) crystals located in the magnetic
volume covering the same solid angle as the charged particle
tracking system.~
The energy resolution for electromagnetic showers is 
$(\sigma_E/E)^2 = (0.013)^2 + (0.0007/E)^2 + (0.008/E^{1/4})^2$, 
where $E$ is in GeV. 

Electron identification is based on a combination of $dE/dx$ measurements
in the CDC, the response of the ACC, and the position, shape 
and the ratio of the total cluster energy registered in the ECL
 and particle momentum.   
The electron identification efficiency, determined by embedding simulated
tracks in multihadron data, is greater than 92\% for tracks reconstructed
in the CDC with $\left|\bm{p}_{\rm lab}\right|>1.0$~GeV/$c$.
The hadron misidentification probability, 
determined using $K^0_S\to \pi^+\pi^-$ decays, is below $0.3\%$.  

The 1.5~T magnetic field is returned via an iron
yoke that is instrumented to detect muons and $K_L$ mesons~(KLM).  
The KLM covers polar angles between $\theta=20^{\circ}$ and $155^{\circ}$ 
and the overall muon identification efficiency, determined by a track embedding
study similar to that used for the electron case, is greater than 87\%
for tracks reconstructed in the CDC 
with $\left|\bm{p}_{\rm lab}\right| > 1$~GeV/$c$. 
The corresponding pion misidentification probability determined
from $K^0_S\to \pi^+\pi^-$ decays is less than 2\%.


\section{Event Reconstruction}

We reconstruct $\bdse$ decays using the decay chain 
$D^{*+} \rt D^0\pi^+$, $D^0 \rt K^-\pi^+$ and  requiring the electron
to be cleanly identified.
In order to suppress the jet-like $e^+e^- \rt q\bar{q}$ continuum background, 
the ratio of the second to zeroth Fox-Wolfram moments~\cite{fw} 
is required to be less than 0.4.
This requirement removes 43\% of the continuum background whilst 
it retains 93\% of the $\bdse$ signal.
Each event is required to contain at least one electron candidate
with momentum in the CM frame in the range 1.00~$\umon$ 
to 2.45~$\umon$.

After a $K^-\pi^+$ combination is selected, 
a $D^0$ decay vertex is reconstructed.
$D^0$ candidates are required to satisfy track quality cuts based on their 
impact parameters relative to the $K^-\pi^+$ vertex:
$\Delta r < 0.2$ cm and $\Delta z < 0.5$ cm.
The $\da$ candidates must have an invariant mass within $3\sigma$ 
of the nominal $D^0$ value, where $\sigma \simeq 6~\umassm$.

\begin{figure}[h]
\centering
\normalsize
\resizebox{8cm}{!}{
\includegraphics{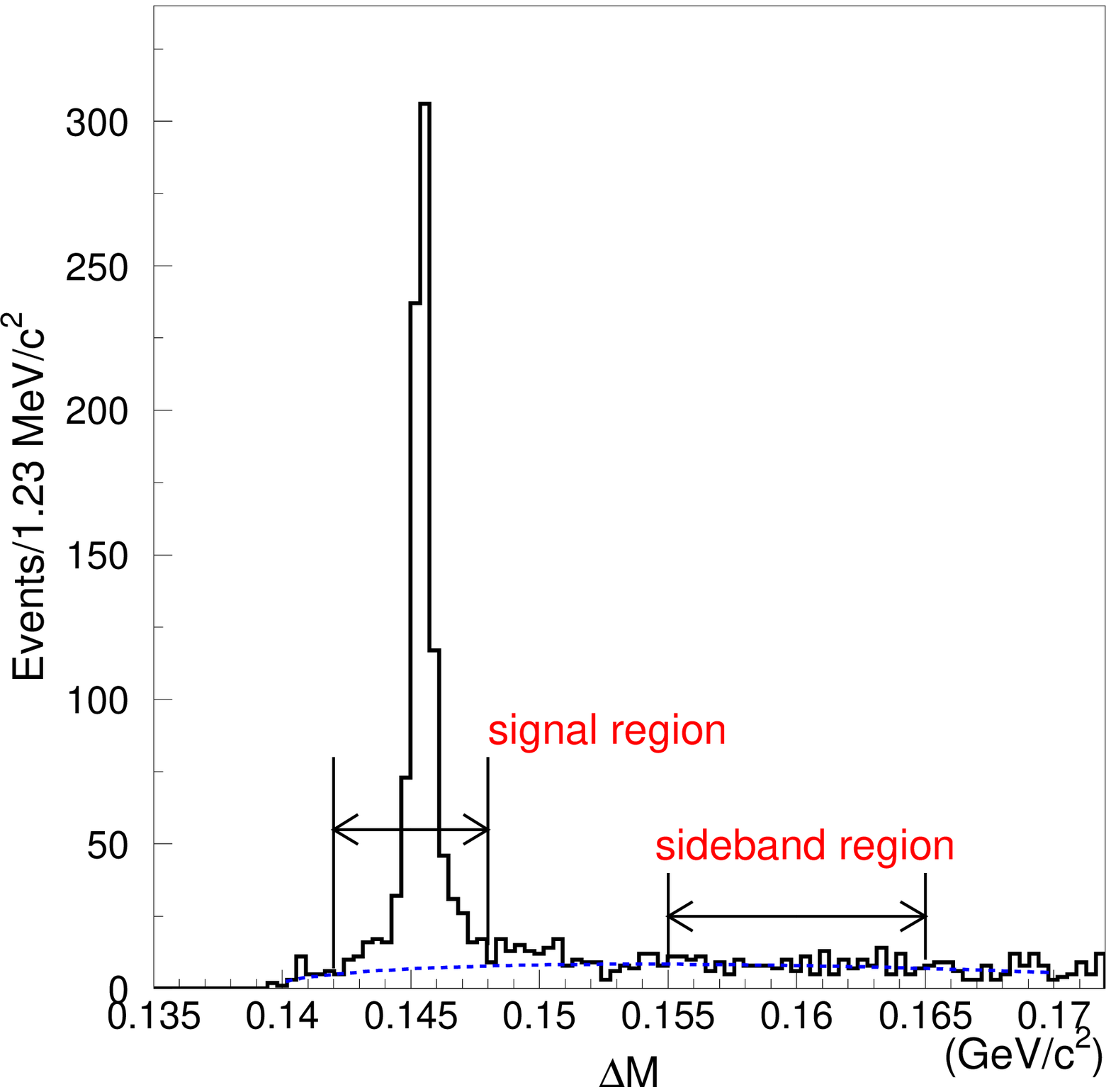}}\\
\caption{
$\Delta M$ distribution for $\bdse$ candidates.
The sideband used to subtract the combinatorial background is 
shown by long arrows and the signal region is indicated by short arrows.
The data (histogram) are superimposed with the combinatorial background 
distribution (dashed curve).
}
\label{fig:bg_comb}
\end{figure}

We combine $D^0$ candidates with slow pion candidates ($\pi^+_s$) 
to fully reconstruct $D^{*+}$ mesons in the mode $\ds$.
The mass difference ($\Delta M = M_{K^-\pi^+\pi^+_s} - M_{K^-\pi^+}$)
is required to lie within 3 $\umassm$ of the nominal peak position 
(Figure~\ref{fig:bg_comb}).
Furthermore the momentum of $\dst$ candidates must satisfy  
$\left|\bm{p}_{\dst}\right| < 0.5\sqrt{E^2_{\rm{beam}}-M^2_{\dst}}$
to be consistent with a $B$ decay hypothesis, where $E_{\rm{beam}}$
is the beam energy.
This requirement retains nearly 100\% of signal events
whilst it rejects 34\% of the continuum background.
A $D^0-e^-$ vertex is constructed and the electron momentum vector
is recomputed with the vertex constraint.

Since the only missing particle in the $\bdse$ decay is the 
neutrino, we expect the square of the missing four-momentum to vanish, 
$p^2_{\rm miss} \equiv p^2_{\bar{\nu}} = (p_{\bar{B^0}} - p_{\dl})^2 = 0$, 
where $p_{\dl} = p_{\dst} + p_{e^-}$. 
This can be expressed in the CM frame: 

\[
p^{2}_{\rm miss} = M^2_{\bar{B^0}} + M^2_{\dl} - 2E_{\bar{B^0}}E_{\dl} + 
2\left|\bm{p}_{\bar{B^0}}\right|\left|\bm{p}_{\dl}\right|
\cos~\theta_{\bar{B^0},\dl},
\]

\noindent
where $E_{\bar{B^0}}$ is replaced by $E_{\rm{beam}}$ 
from which $\left| \bm{p}_{\bar{B^0}}\right|$ is calculated. 
Neglecting the last term since 
$\left| \bm{p}_{\bar{B^0}}\right|$ is small in the CM frame 
($\sim 340~\umonm$), we define the square of missing mass as follows: 

\[
M^2_{\rm miss} =  M^2_{\bar{B^0}} + M^2_{\dl} - 2E_{\bar{B^0}}E_{\dl}.
\]

\noindent
We make the cut $M^2_{\rm miss} < 1$~$\umasss$ to suppress the background from
$\bar{B^0} \rt D^{**}e^-\bar{\nu}$, which has an additional pion
(by $D^{**}$, we denote the sum of resonant and non-resonant states).
The efficiency of the cut on $M^2_{\rm miss}$ is about 96\% for 
$\bdse$ and 53\% for $\bar{B^0} \rt D^{**}e^-\bar{\nu}$.
Using the condition $p^2_{\rm miss} = 0$,
we can extract $\cos\theta_{\bar{B^0},\dl}$ as follows:

\[
\cos\theta_{\bar{B^0},\dl} = \frac{ 2E_{\bar{B^0}}E_{\dl} - 
M^2_{\bar{B^0}} - M^2_{\dl}}
{2\left| \bm{p}_{\bar{B^0}}\right| \left| \bm{p}_{\dl} \right|}.
\]

\noindent
The quantity $\cos \theta_{\bar{B^0},\dl}$ is highly correlated 
with the estimated value of $M^2_{\rm miss}$ but allows us to 
impose a kinematic consistency condition 
through the requirement $\left|\cos\theta_{\bar{B^0},\dl}\right| < 1$.
This requirement removes 66\% of the
$\bar{B^0} \rt D^{**}e^-\bar{\nu}$ background and 81\% of the continuum 
background whilst keeping 88\% of the signal. 

The reconstructed $\bar{B^0}$ mass $M_{\bar{B^0}}^{\rm rec}$ is defined as

\[
M_{\bar{B^0}}^{\rm rec} = \sqrt{ E^2_{\rm beam} - \left| \bm{p}_{\dst} + 
                               \bm{p}_{e^-} + 
                               \bm{p}_{\bar{\nu}} \right|^2 },
\]

\noindent
where the neutrino momentum is calculated using all the reconstructed tracks
and clusters in the event: ${\bm{p}}_{\bar{\nu}} = -\sum_{i} \bm{p}_i$.
In this analysis, we use this information only to make a loose 
consistency requirement of $M_{\bar{B^0}}^{\rm rec} >$ 5.0~$\umass$.
This cut value gives 95\% efficiency for the 
signal mode; it removes 18\% of continuum background. 

Since this analysis is not based on a full reconstruction of the $\bar{B}^0$,
we do not calculate $q^2$ using missing momentum information.
Instead, we compute ${\rm y}$ by taking the average of the values obtained 
with the two extreme configurations of the $\bar{B^0}$ and $\dst$ directions
consistent with $\bm{p}_{D^{*+}},~\bm{p}_{e^{-}}$ and the value of
$\cos \theta_{\bar{B^0},\dl}$ calculated above.
We reject the unphysical region in the distribution of $\tilde{\rm y}$,
which denotes the measured ${\rm y}$ value.
The resolution of ${\rm y}$ has been studied using MC, and is found 
to be accurately modeled by a symmetric Gaussian with $\sigma \simeq 0.04$.

The overall signal reconstruction efficiency is 6.0\% after all event
selection criteria.


\section{Background Subtraction}

The background sources fall into five categories:
combinatorial, correlated, uncorrelated, fake electron and continuum.
The largest source is the combinatorial background in the 
$D^{*+}$ reconstruction.
The level of this background is determined from a fit to the 
$\Delta M$ distribution, using a phase space shape for the background.
The corresponding shape of the background in the variable $\tilde{\rm y}$
is deduced from events in a sideband region, which is defined as 0.155 $\umass 
< \Delta M <$ 0.165 $\umass$ as shown in Figure~\ref{fig:bg_comb}.

The contributions from other sources of background are estimated
using MC. 
Correlated background occurs when a $\bar{B^0}$ meson decays to a 
final state containing a $D^{*+}$ and an $e^-$ through channels 
other than $\bdse$.
The largest source is due to the process 
$\bar{B^0} \rt D^{**} e^- \bar{\nu}$. 
From a simulation based on the ISGW2~\cite{isgw2}, and Goity and Roberts
models~\cite{goity},
we find this background accounts for about 9\% of the raw yield.
Hence this is a large source of background.

In uncorrelated background events, the $D^*$ originates
from the decay of the $\bar{B}$ and the electron originates from the 
partner $B$.
The electrons in this background must come from secondary $B$ meson
decay products or $B^0-\bar{B^0}$ mixing
in order to have the correct charge.
It is also possible to have electrons from the decay or misidentification
of light hadrons and these are classified as fake electron background. 
The fake electron background is estimated using MC without primary electrons 
from $\bar{B}$ decays. A very small contribution to the signal yield is found.
The size of the continuum background is measured using $q\bar{q}$ MC.

\begin{figure}[h]
\centering
\resizebox{16cm}{!}{
\includegraphics{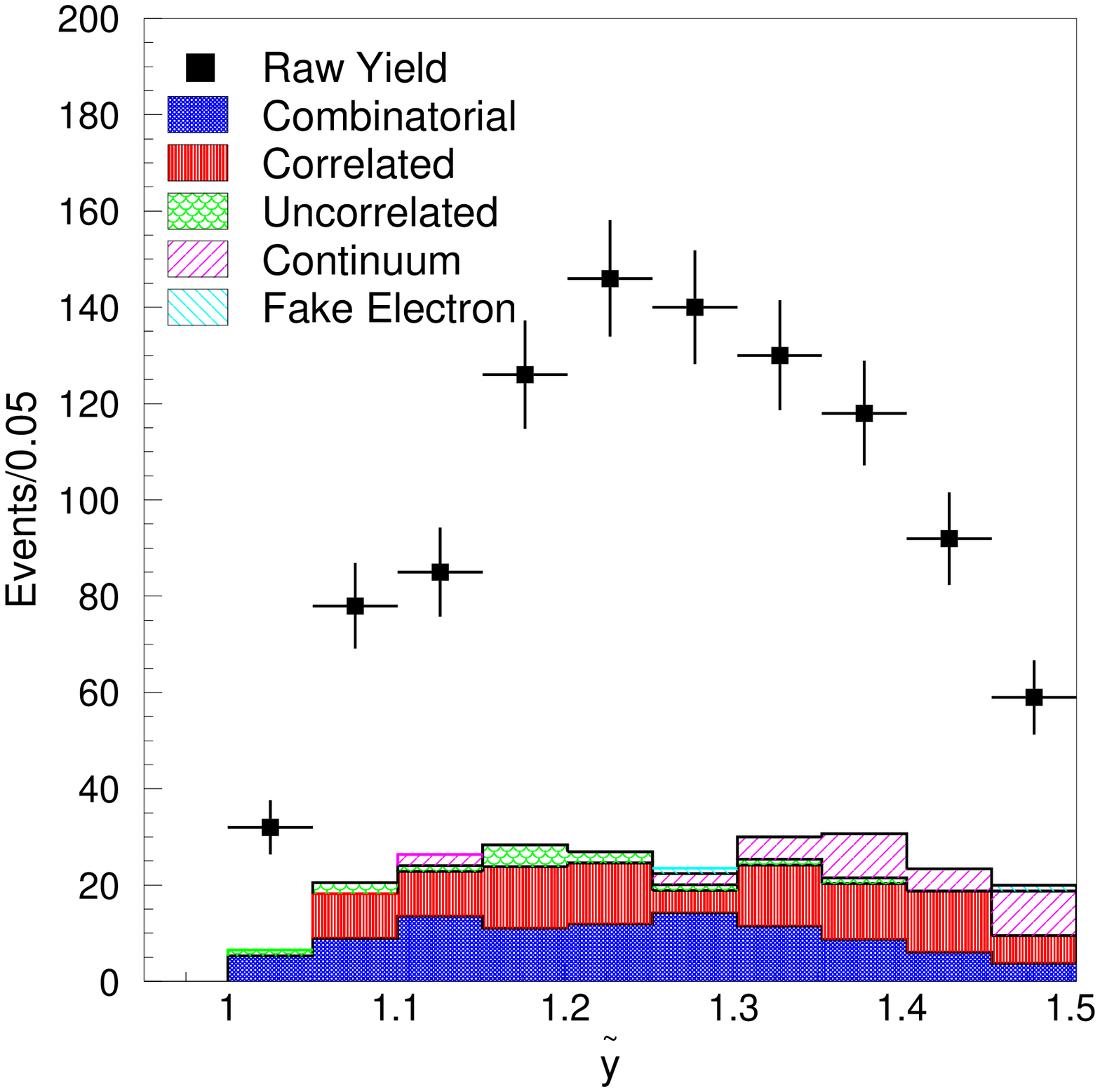}
\includegraphics{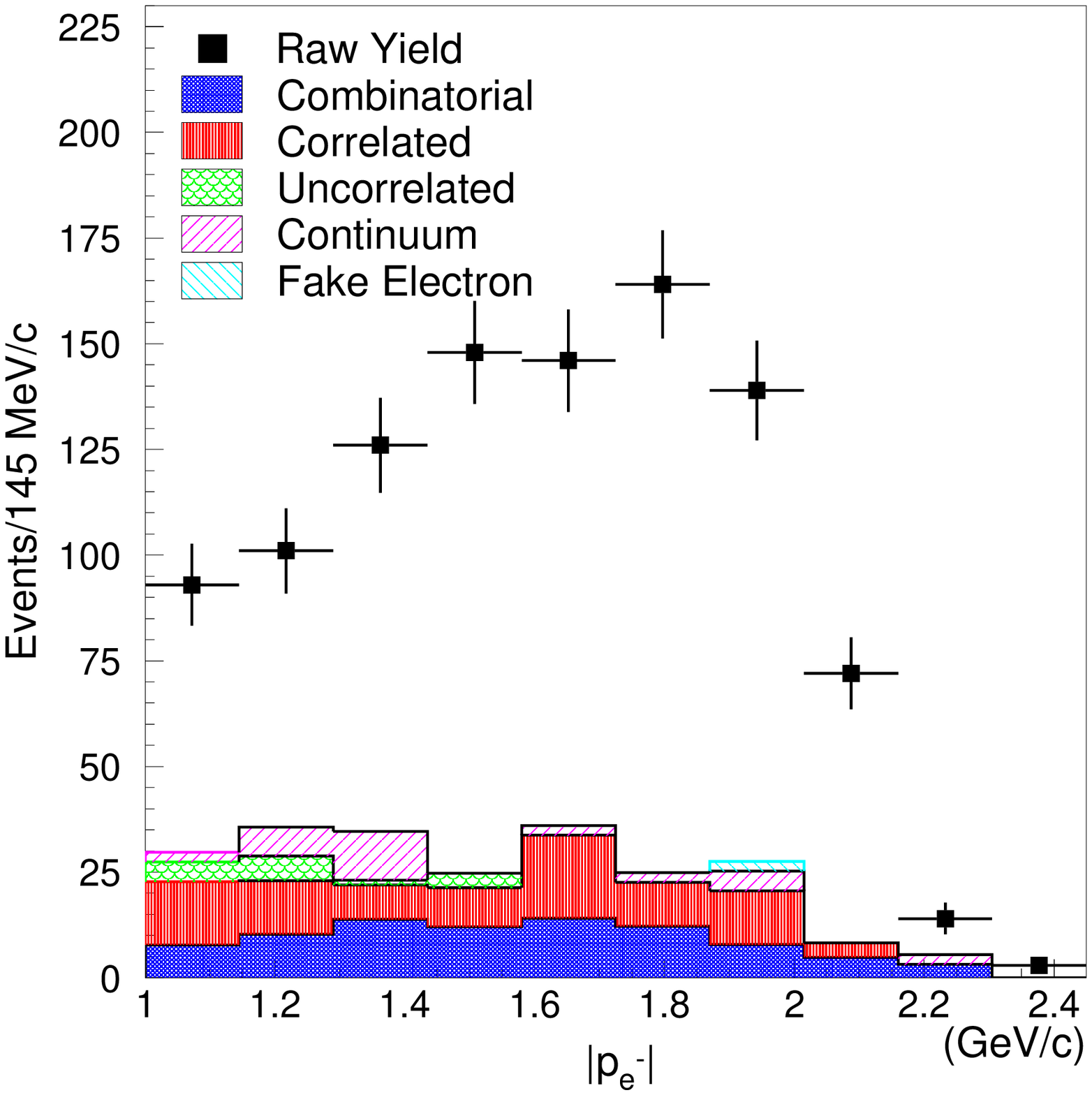}}\\
\caption{
Left (right) plot shows the raw yield and background 
as a function of $\tilde{\rm y}$ (electron momentum in the CM frame). 
From the bottom, combinatorial, correlated, 
uncorrelated, continuum and fake electron backgrounds are shown.
}
\label{fig:bg}
\end{figure}

In Figure~\ref{fig:bg}, the raw yield and various backgrounds are 
displayed as a function of $\tilde{\rm y}$ and electron momentum in the CM 
frame. Table~\ref{tbl:bg_sub} shows the expected background contributions to 
the raw yield.


\section{Form Factor Parameterization}

The differential rate for the $\bdse$ decay is given by~\cite{diff_rate}

\[
\frac{d\Gamma}{d{\rm y}}=\frac{G_F^2}{48\pi^3}
 M^3_{D^{*+}}(M_{\bar{B^0}} - M_{D^{*+}})^2 g({\rm y}) |V_{cb}|^2 F({\rm y})^2,
\]

\[
 g({\rm y})  = 
\sqrt{{\rm y}^2-1}({\rm y}+1)^2 \left[1+4\frac{{\rm y}}{{\rm y}+1}
\frac{1-2{\rm y}r+r^2}{(1-r)^2}\right],
\]

\noindent
where $G_F$ is the Fermi coupling constant and $r = M_{D^{*+}}/M_{\bar{B^0}}$. 
We approximate $F({\rm y})$ with a Taylor series expansion around y = 1:

\[
 F({\rm y}) = F(1)[1 - \rhf({\rm y}-1) + c({\rm y}-1)^2 +
 \mathcal{O}({\rm y}-1)^3]. 
\]

\noindent
Early $\vcb$ measurements~\cite{vcb_argus}-\cite{vcb_delphi1} 
used this expansion neglecting the second order term 
(this is called the linear form factor parameterization).
However, since a determination of $\vcb$ requires the 
extrapolation of the differential decay rate to y = 1,
constraints on the shape of the form factor are highly desirable
to reduce the uncertainties associated with this extrapolation.
A suitable framework to derive such constraints on heavy meson form factors
has recently been provided by dispersion 
relations~\cite{disper_1992}-\cite{disper_1997}.

To use the relevant dispersion relation~\cite{disper_1998}
we relate $F({\rm y})$ to the axial vector form factor $h_{A_1}({\rm y})$ 
using~\cite{HQET,Neubert_FA1}

\[
\begin{array}{lc}
 g({\rm y})F({\rm y})^2 = \sqrt{{\rm y}^2-1}({\rm y}+1)^2  h_{A_1}({\rm y})^2 
\\ \nonumber
 \times \left\{2\left[\frac{1-2{\rm y}r+r^2}{(1-r)^2}\right]
 (1+ R_1({\rm y})^2\frac{{\rm y}-1}{{\rm y}+1})  
 + \left[ 1+ (1-R_2({\rm y}))\frac{{\rm y}-1}{1-r} \right]^2\right\}.
\end{array}
\]

\noindent
$R_1({\rm y})$ and $R_2({\rm y})$ are given by:

\[
\begin{array}{lc}
R_1({\rm y}) \approx  R_1(1) - 0.12({\rm y}-1) + 0.05({\rm y}-1)^2 , 
\\ \nonumber
R_2({\rm y}) \approx  R_2(1) + 0.11({\rm y}-1) - 0.06({\rm y}-1)^2.
\end{array}
\]

\noindent
Using the results of the dispersion relation, 
$h_{A_1}({\rm y})/h_{A_1}(1)$ depends only on a single unknown parameter 
$\rha$:

\[
\frac{h_{A_1}({\rm y})}{h_{A_1}(1)} \approx 
 1-8\rha z + (53\rha - 15)z^2 - (231\rha- 91)z^3 ,
\]

\noindent
where $z = (\sqrt{{\rm y}+1} - \sqrt{2})/(\sqrt{{\rm y}+1} + \sqrt{2})$.
It can be seen that $h_{A_1}({\rm y}) \rt F({\rm y})$ in the limit y $\rt$ 1, 
hence $h_{A_1}(1) = F(1)$.
Our main analysis uses this dispersion relation parameterization.
The dispersive bound used in this analysis requires that $-0.14 < \rha < 1.54$
~\cite{disper_1998}.


\section{Fit and Results}

\begin{figure}[b]
\centering
\normalsize
\resizebox{8cm}{!}{
\includegraphics{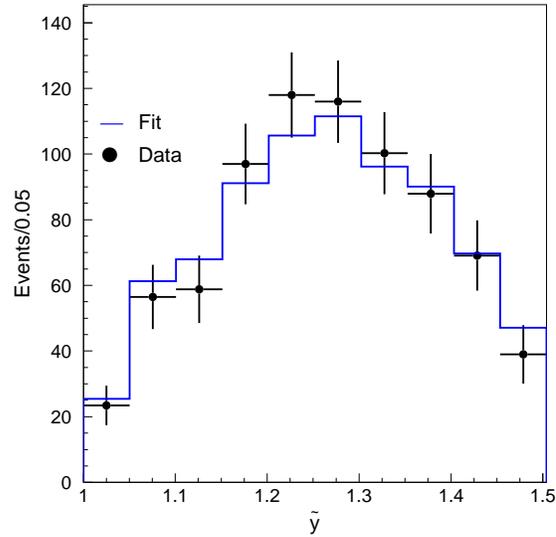}}\\
\caption{ Yield of $\bdse$ as a function of $\tilde{\rm y}$. 
	 The points are background subtracted data and
         the histogram represents the fit to data.}
\label{fig:yfit_chi2}
\end{figure}

After subtraction of background,
we fit the final yield as a function of y for $|V_{cb}|F(1)$ and $\rha$. 
We use $R_1(1) = 1.3\pm0.1$ and $R_2(1) = 0.8\pm0.2$
which are obtained  using HQET and QCD sum rules~\cite{HQET}.
We minimize the following variable:

\[
 \chi^2  = \sum_{i=1}^{N_{\rm{bin}}=10}\left(\frac{ N_{i}^{\rm{obs}} - 
           \sum_{j=1}^{N_{\rm{bin}}} 
           {\epsilon_{ij} N_j}}  {\sigma_{N_{i}^{\rm{obs}}}} \right)^2,
\]

\noindent
where $N_{\rm{bin}}$ is the number of bins over the range 
1.000 $< \tilde{\rm y} <$ 1.504,
$N^{\rm{obs}}_i$ is the yield in the $i$th $\tilde{\rm y}$ bin,
$\sigma_{N_{i}}^{\rm{obs}}$ is the statistical error of $N^{\rm{obs}}_i$ and 
$N_j$ is the number of decays in the $j$th bin, given by

\[
  N_j =  2 N_{B\bar{B}} f_{00} \tau_{\bar{B^0}} {\mathcal{B}}(\ds) 
         {\mathcal{B}}(\da) \int_{\hspace{0.5cm} {\rm y}_j}
         \frac{d\Gamma}{d{\rm y}} d{\rm y},
\]
\noindent 
where the integral is over the range of y in the $j$th bin, 
$N_{B\bar{B}}$ is the number of $B\bar{B}$ pairs in the sample and
$f_{00}$ is the $\Upsilon(4S) \rt B^0\bar{B}^0$ branching fraction 
and is taken to be 0.5;
$\tau_{\bar{B^0}}$ is the $\bar{B^0}$ lifetime, 
${\mathcal{B}}(\ds)$ is the $\ds$ branching fraction, 
${\mathcal{B}}(\da)$ is the $\da$ branching fraction and
these values are taken from~\cite{pdg}.
The efficiency matrix element $\epsilon_{ij}$ gives the probability that 
events generated in bin $j$ in y are reconstructed in bin $i$.
Typical values of the diagonal elements are in the range 0.023$-$0.033, 
while the elements for neighboring bins are in the range 0.007$-$0.021.
A result of

\[
\begin{array}{lc}
\vcb F(1) = \vfst, \\ \nonumber
\rha = \rhst \nonumber
\end{array}
\]

\noindent
is obtained, where the errors are statistical only.
The correlation coefficient between $\vcb F(1)$ and $\rha$ is 0.91
and the $\chi^2$ of the fit is 3.38 for 8 degrees of freedom.
The fit results are shown in Figure~\ref{fig:yfit_chi2}.
By summation of the $N_j$ obtained above over all bins, the 
branching fraction of ${\mathcal{B}}(\bdse)$ is determined to be

\[
 {\mathcal{B}}(\bdse) = (\brst) \times 10^{-2},
\]

\noindent
where again the error is statistical only.

We have studied the impact of precise values of the 
$R_1(1)$ and $R_2(1)$ coefficients on the results of our analysis by
using the CLEO measurement, 
$R_1(1) = 1.18\pm0.32$ and $R_2(1) = 0.71\pm0.23$~\cite{cleo_R1R2}.
In addition, we have studied the case of the linear form factor defined
above. This is a common approximation in the limit of heavy quark 
symmetry, in which $R_1({\rm y}) = R_2({\rm y}) = 1 $~\cite{HQET}.
In Table~\ref{tbl:vcbfit} the results of the three different fits are  
compared.

\begin{figure}[t]
\centering
\normalsize
\resizebox{8cm}{!}{
\includegraphics{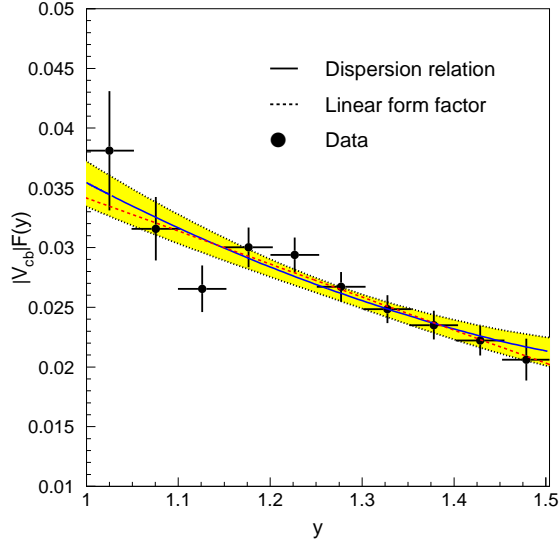}}\\
\caption{$\vcb F({\rm y})$ as a function of y. 
The data points with errors are derived from unfolding.
The curves show the result obtained using the Caprini {\it et~al.} form factor 
constrained by a dispersion relation (solid) and a linear form factor (dashed).
The statistical uncertainty of the Caprini {\it et~al.} form factor fit is
shown by the shaded band.} 
\label{fig:yfit_unfd}
\end{figure}

In order to deal with the smearing effects due to the experimental
resolution, an unfolding method is applied~\cite{unfolding}. 
The unfolded data of $\vcb F({\rm y})$ as a function of y
is shown in Figure~\ref{fig:yfit_unfd}.


\section{Systematic Errors}

The systematic uncertainties in the result from each possible source
are estimated by changing the input values in the $\vcb$ fitting procedure. 
A summary of the contributions to the
systematic errors is given in Table~\ref{tbl:syst}.

The systematic uncertainty in the number of $B\bar{B}$ events is 
estimated by varying $N_{B\bar{B}}$ within its measured uncertainty.
The various fit inputs, the $\bar{B^0}$ lifetime and $D^{*+}, D^0$
branching fractions, are varied according to the errors given in~\cite{pdg}.
Since the calculation of the branching fraction in~\cite{pdg} is based on
the assumption of equal $B^+B^-$ and $B^0\bar{B^0}$ production, 
we do not take into account the uncertainty of $f_{00}$.

The uncertainty in the tracking efficiency for high momentum tracks
is determined using $\eta \rt \pi^+\pi^-\pi^0$ and 
$\eta \rt \gamma\gamma$ decays.
By comparing the numbers of reconstructed $\eta$'s between data and MC,
we obtain an error of 2.5\% on $\vcb F(1)$.

The efficiency for identifying electrons is evaluated by embedding
single simulated electrons into hadronic events as well as
$J/\psi \rt e^+e^-$ events. 
We find that the uncertainty in the electron identification efficiency leads
to a 1.0\% systematic error on $\vcb F(1)$.

For $D^0$ reconstruction, three efficiencies are combined: kaon 
identification, pion identification and vertexing efficiencies.
The particle identification efficiencies are obtained from data using the 
fact that the slow pion from $D^{*+}$ decay allows the decay,
$D^{*+} \rt D^0\pi^+$, followed by $D^0 \rt K^-\pi^+$, to be reconstructed
without relying on particle identification.
The systematic error in $D^0$ reconstruction is calculated from 
the measurement error (2.2\%) of the efficiency in data.
We find a systematic error of 1.4\% on $\vcb F(1)$.

The slow pion reconstruction is mainly affected by the event environment
and by material in the detectors.
The former effect is evaluated by the track embedding method.
It results in an uncertainty of 3.4\%.
The latter effect is estimated to be 3.7\% by varying the energy loss
in the inner detectors in the MC simulation.
In total we find an uncertainty of 5.0\% in the slow pion efficiency, leading 
to 2.6\% uncertainty in $\vcb F(1)$.

The fit of $\vcb F(1)$ depends on the resolution of y, 
which is a function of the $\bar{B^0}$ and $D^{*+}$ four-momentum vectors 
and of the beam energy. Therefore we change the $\bar{B^0}$ mass,
$D^{*+}$ mass and the beam energy by their uncertainties 
to estimate the contribution from these sources.
We find a 1.1\% change in $\vcb F(1)$.

We estimate the uncertainty from the combinatorial background subtraction
by changing the sideband region and the fitting function,
and assign a systematic error of 1.0\% on $\vcb F(1)$.

By varying the branching fractions of the contributing modes in MC,
the systematic error due to correlated background is determined to be 1.4\%.
We also estimate the size of the $\bar{B^0} \rt D^{**}e^-\bar{\nu}$ 
background from data by fitting the $M_{\rm{miss}}^2$ distribution.
In this fit, the normalizations of the other background components
are fixed and the shapes of the $D^{*+}e^-\bar{\nu}$
and $D^{**}e^-\bar{\nu}$ events come from MC.
Using this method, we find $72\pm8$ background events, which is compatible
with the number derived from MC within the systematic error.

The fraction of uncorrelated background in MC
is varied by 100\%, leading to a systematic error of 0.4\% on $\vcb F(1)$.
The uncertainty due to the fake electron background is estimated by varying
the misidentification rate in MC according to the measured electron 
misidentification rate  0.20$\pm$0.02\% determined from 
$K^0_S \rt \pi^+\pi^-$ decays.
The continuum background may be miscalculated if the $q\bar{q}$ background 
is inaccurately modeled in MC.
The size of the continuum background is varied by 50\%,
giving 0.5\% uncertainty on $\vcb F(1)$.
We also check this background using the 0.6~$\fb$ off-resonance data sample;
the result is compatible with the MC estimate.

The uncertainty related to the finite MC statistics used to measure the 
efficiency matrix is measured by including the errors of the matrix
elements in the fit.
We denote this effect, 1.7\% on $\vcb F(1)$, as MC statistics.

The $R_1(1)$ and $R_2(1)$ are related to the 
electron momentum distribution.
To estimate this effect, we vary  $R_1(1)$ and $R_2(1)$
within the errors given by~\cite{HQET} and find a resultant 
11.8\% uncertainty on $\rha$.

To test the stability of the result, several systematic checks were made.
The requirements on the second Fox-Wolfram moment, the $K^-\pi^+$ invariant
mass and the square of the missing mass were
individually tightened; in each case no discrepancy was found.
The analysis was also performed for a narrower range of electron
momentum; again a consistent result was found.


\section{Conclusion}

In conclusion, 
we have measured the y distribution of $\bdse$ decays to obtain 
$\vcb F(1)$ and the slope of the form factor using a dispersion relation.
The results are 

\[
\vcb F(1) = \vf, 
\]

\[
\rha = \rh,
\]

\noindent
with a correlation coefficient between $\vcb F(1)$ and $\rha$ of 0.91.
From the same fit, we extract the branching fraction

\[
{\mathcal{B}}(\bdse) = \br.
\]

\noindent
These results are consistent with other published measurements
from CESR~\cite{vcb_cleo1} and 
LEP~\cite{vcb_opal,vcb_delphi2}.

Using $F(1)$ = 0.913$\pm$0.042~\cite{Babar}, 
the following value of $\vcb$ is obtained:

\[
\left| V_{cb} \right| = \vcbf,
\]

\noindent
where the first error is 
statistical, the second is systematic and the third is theoretical.


{\bf Acknowledgments}\\

We wish to thank the KEKB accelerator group for the excellent
operation of the KEKB accelerator. We acknowledge support from the
Ministry of Education, Culture, Sports, Science, and Technology of Japan
and the Japan Society for the Promotion of Science; the Australian
Research
Council and the Australian Department of Industry, Science and
Resources; the Department of Science and Technology of India; the BK21
program of the Ministry of Education of Korea and the Center for High
Energy Physics sponsored by the KOSEF; the Polish
State Committee for Scientific Research under contract No.2P03B 17017;
the Ministry of Science and Technology of Russian Federation; the
National Science Council and the Ministry of Education of Taiwan; and
the U.S. Department of Energy.


\newpage

\begin{table}[b]
\centering
\caption{ Yield of the $\bdse$ and the estimated background
contributions. The errors are statistical only.}
\begin{tabular}{l|c}\hline 
{Raw Yield}&{1006$\pm$32}\\
\hline
{Combinatorial background}&{100$\pm$7}\\
{Correlated background}&{91$\pm$10}\\
{Uncorrelated background}&{15$\pm$4}\\
{Fake electron background}&{2$\pm$2}\\
{Continuum background}&{32$\pm$6}\\
\hline
{Final Yield}&{766$\pm$35}\\
\hline
\end{tabular}
\label{tbl:bg_sub}
\end{table}

\newpage

\vspace{1cm}

\newpage

\vspace{1cm}

\begin{table}[b]
\centering
\caption{Summary of fit results according to different form factor (FF) 
parameterizations, where the errors are statistical only.
Our main analysis uses values of $R_1(1)$, $R_2(1)$ from QCD sum rules.}
\hspace{-2cm}

\begin{tabular}{lccccc} \hline
{FF shape~\&~$R_1(1),R_2(1)$}&{$|V_{cb}|F(1)\cdot10^{2}$}&{$\rha$}&
{$\rhf$}&{$\mathcal{B}(\bdse)$}&{$\chi^2/{\rm{ndf}}$}\\
\hline
{Dispersive~\&~QCD sum rules}&
{3.54$\pm$0.19}&{1.35$\pm$0.17}&{$\cdot$}&{(4.59$\pm$0.23)\%}&{3.38/8}\\
{Dispersive~\&~CLEO value}&
{3.58$\pm$0.19}&{1.45$\pm$0.16}&{$\cdot$}&{(4.60$\pm$0.23)\%}&{3.79/8}\\
{Linear~\&~Heavy quark limit}&
{3.42$\pm$0.17}&{$\cdot$}&{0.81$\pm$0.12}&{(4.57$\pm$0.24)\%}&{2.23/8}\\
\hline
\end{tabular}
\label{tbl:vcbfit}
\end{table}

\newpage
\begin{table*}[h]
\caption{Summary of systematic errors. For each possible source the error on each of the three measured values is given.}
\begin{tabular}{lccc}
\hline
{Error Sources}&
{$|V_{cb}|F(1)$ (\%)}&{$\rha$ (\%)}&{${\mathcal{B}}(\bdse)$~(\%)}\\
\hline
{$N_{B\bar{B}}$}&{0.5}&{$\cdot$}&{1.0}\\
{$\tau_{\bar{B^0}}$}&{1.0}&{$\cdot$}&{$\cdot$}\\
{${\mathcal{B}}(\ds)$}&{0.4}&{$\cdot$}&{0.7}\\
{${\mathcal{B}}(\da)$}&{1.2}&{$\cdot$}&{2.3}\\
{Tracking efficiency}&{2.5}&{$\cdot$}&{5.1}\\
{Electron ID efficiency}&{1.0}&{$\cdot$}&{1.9}\\
{$D^0$ reconstruction efficiency}&{1.4}&{$\cdot$}&{2.8}\\
{Slow pion efficiency}&{2.6}&{1.2}&{5.0}\\
\hline
{Subtotal}&{4.3}&{1.2}&{8.3}\\
\hline
{y resolution}&{1.1}&{2.8}&{0.5}\\
{Combinatorial BG}&{1.0}&{0.1}&{1.9}\\
{Correlated BG~($D^{**}$)}&{1.4}&{4.0}&{0.5}\\
{Uncorrelated BG}&{0.4}&{0.5}&{0.5}\\
{Fake electron BG}&{$\cdot$}&{0.2}&{0.1}\\
{Continuum BG}&{0.5}&{2.9}&{0.9}\\
{MC statistics}&{1.7}&{4.4}&{1.3}\\
{$R_1(1), R_2(1)$}&{1.1}&{11.8}&{0.3}\\
\hline
{Subtotal}&{2.9}&{13.7}&{2.6}\\
\hline
{Total}&{\bf{5.2}}&{\bf{13.8}}&{\bf{8.7}}\\
\hline
\end{tabular}
\label{tbl:syst}
\end{table*}


\begin{thebibliography}{99}

\bibitem{Wolf} L.~Wolfenstein, Phys. Rev. Lett. {\bf 51}, 1945 (1983). 

\bibitem{HQET} M.~Neubert, Phys. Rep.~{\bf 245}, 259 (1994). 

\bibitem{disper_1998} I.~Caprini, L.~Lellouch, M.~Neubert, Nucl. Phys. 
B {\bf 530}, 153 (1998).

\bibitem{sirlin} A.~Sirlin, Nucl. Phys. B {\bf 196}, 83 (1982).

\bibitem{luke} M.~Luke, Phys. Lett. B {\bf 252}, 447 (1990).

\bibitem{dlnu_mq} M.~Neubert and V.~Rieckert, Nucl. Phys. 
B {\bf 382}, 97 (1992).

\bibitem{dlnu1} D.~Buskulic {\it et~al.} [ALEPH Collaboration], Phys. Lett.
B {\bf 395}, 373 (1997).

\bibitem{dlnu2} J.~Bartelt {\it et~al.} [CLEO Collaboration], Phys. Rev. Lett.
{\bf 82}, 3746 (1999).

\bibitem{dlnu_belle} K. Abe {\it et~al.} [Belle Collaboration], 
hep-ex/0111082, submitted to Phys. Lett. B.

\bibitem{Babar} M. Neubert in BaBar Physics Book, P.~F.~Harrison and 
H.~R. Quinn (editors), SLAC-R-504 (1998).

\bibitem{lattice_F1} S.~Hashimoto {\it et~al.}, hep-ph/0110253,
FERMILAB-PUB-01/317-T (2001).

\bibitem{vcb_opal} G.~Abbiendi {\it et~al.} [OPAL Collaboration], Phys. Lett. 
B {\bf 842}, 15 (2000).

\bibitem{vcb_delphi2} P. Abreu {\it et~al.} [DELPHI Collaboration], Phys. Lett. 
B {\bf 510}, 55 (2001).

\bibitem{vcb_cleo2} J.~P.~Alexander {\it et~al.} [CLEO Collaboration], 
hep-ex/0007052, CLEO-CONF-00-03 (2000).

\bibitem{NIM}{K. Abe {\it et al.} [Belle Collaboration], 
	KEK Progress Report 2000-4 (2000),
	to be published in Nucl. Inst. and Meth. A.}

\bibitem{kekb}{KEKB B Factory Design Report, KEK Report 95-7 (1995),
	unpublished; Y. Funakoshi {\it et al.}, Proc. 2000
        European Particle Accelerator Conference, Vienna (2000).}

\bibitem{geant}{R. Brun {\it et al.}, GEANT 3.21, CERN Report No.
        DD/EE/84-1 (1987).}
        
\bibitem{fw} G.~Fox and S.~Wolfram, Phys. Rev. Lett {\bf 41}, 1581 (1978). 

\bibitem{isgw2} N.~Isgur and D.~Scora, Phys. Rev. D {\bf 52}, 2783 (1995);
N.~Isgur, D.~Scora, B.~Grinstein, and M.~B.~Wise, Phys. Rev. D {\bf 39}, 799 
(1989).

\bibitem{goity} J.~L.~Goity and W.~Roberts, Phys. Rev. D {\bf 51}, 3459 (1995).

\bibitem{diff_rate} N.~Neubert, Phys. Lett. 
B {\bf 264}, 455 (1991); B {\bf 338}, 84 (1994).

\bibitem{vcb_argus} H.~Albrecht {\it et~al.} [ARGUS Collaboration], Z. Phys. 
C {\bf 57}, 533 (1993). 

\bibitem{vcb_cleo1} B.~Barish {\it et~al.} [CLEO Collaboration], Phys. Rev. 
D {\bf 51}, 1014 (1995).

\bibitem{vcb_aleph1} D.~Buskulic {\it et~al.} [ALEPH Collaboration], Phys. Lett.
B {\bf 359}, 373 (1995).

\bibitem{vcb_delphi1} P.~Abreu {\it et~al.} [DELPHI Collaboration], Z. Phys. 
C {\bf 71}, 539 (1996).

\bibitem{disper_1992} E.~de~Rafael and J.~Taron, Phys. Lett.  
B {\bf 282}, 215 (1992).

\bibitem{disper_1994} I.~Caprini, Z. Phys. 
C {\bf 61}, 651 (1994).

\bibitem{disper_1995} C.~G.~Boyd, B.~Grinstein and R.~F.~Lebed, Phys. Lett. 
B {\bf 353}, 306 (1995); Nucl. Phys. B {\bf 461}, 493 (1996).

\bibitem{disper_1996} I.~Caprini and M.~Neubert, Phys. Lett. 
B {\bf 380}, 376, (1996).

\bibitem{disper_1997} C.~G.~Boyd, B.~Grinstein and R.~F.~Lebed, Phys. Rev. 
D {\bf 56}, 6895 (1997).

\bibitem{Neubert_FA1} Int.~J.~Mod.~Phys. A {\bf 11}, 4173 (1996).

\bibitem{pdg}
D.~E.~Groom {\it et al.} [Particle Data Group Collaboration],
Eur.\ Phys.\ J.\ C {\bf 15}, 1 (2000).

\bibitem{cleo_R1R2} J.~E.~Duboscq {\it et~al.} [CLEO Collaboration], Phys. Rev. Lett. 
{\bf 76}, 3898 (1996).

\bibitem{unfolding} G.~D'Agostini, Nucl. Inst. and Meth. A {\bf 362}, 
487 (1995).



\end{thebibliography}
\end{document}